\documentstyle[12pt]{article}
\def\hf{\frac{1}{2}}
\def\thf{\frac{3}{2}}
\def\twotrd{\frac{2}{3}}
\def\expfac{e^{-q^2/6\alpha^2}}
\def\bolde{\mbox{\boldmath $e$}}
\def\bp{\mbox{\boldmath $p$}}
\def\bq{\mbox{\boldmath $q$}}
\def\bx{\mbox{\boldmath $x$}}
\def\bP{\mbox{\boldmath $P$}}
\def\bJ{\mbox{\boldmath $J$}}
\def\bi{\mbox{$i$}}
\def\bigcdot{\mbox{\large $\cdot$\normalsize}}
\def\bigrho{\mbox{\large $\rho$\normalsize}}
\def\bigphi{\mbox{\large $\phi$\normalsize}}
\def\be{\mbox{\Large $e$\normalsize}}

\def\cbvy{\mbox{\large ${\cal Y}$\normalsize}}
\def\bsigma{\mbox{\boldmath $\sigma$}}
\def\bSigma{\mbox{\boldmath $\Sigma$}}
\def\brho{\mbox{\boldmath $\rho$}}
\def\blambda{\mbox{\boldmath $\lambda$}}

\def\bra{\langle }
\def\ket{\rangle }
\def\sumX{\!\!\!\!\!\!\sum_{X=N^*,\Delta^*}} 

\def\Xa{N^*(\hf^-,1535)} 
\def\Xb{N^*(\hf^-,1650)}
\def\Xc{N^*(\thf^-,1520)}
\def\Xd{N^*(\thf^-,1700)}
\def\Xe{\Delta(\thf^+,1232)}
\def\Xf{\Delta^*(\hf^-,1620)}
\def\Xg{\Delta^*(\thf^-,1700)}
 
\def\dEx{\,E(q)-E_X(q)}
\def\dMx{\,M-M_X}
\newcommand\bkk[6]{\langle {#1}\,{#2}\,{#3}\,{#4}\!\mid\!{#5}\,{#6} \rangle}

\newcommand\ca[2]{\,\bra\chi^{#1}_{_{#2}}|\bsigma_3|\chi^{\lambda}_{_{m_N}}\ket\,}
\newcommand\cb[1]{\,\bra\chi^{\lambda}_{_{#1}}|\bsigma_3|\chi^{\lambda}_{_{m_N}}\ket\,}

\newcommand\trija[6]{(-1)^{\hf+{#6}} \sqrt{2{#5}\!+\!1} 
\left(\begin{array}{ccc}\!\!{#1} &{#3} &{#5}\\ 
                         \!\!{#2} & {#4} & -{#6}\end{array}\!\!\right) }
\newcommand\trijb[6]{(-1)^{-\hf+{#6}} \sqrt{2{#5}\!+\!1} 
\left(\begin{array}{ccc}\!\!{#1} &{#3} &{#5}\\ 
                         \!\!{#2} & {#4} & -{#6}\end{array}\!\!\right) }

\begin{document}  
\begin{flushright}
ADP-96-15/T218 \\
(to appear in the Australian Journal of Physics)
\end{flushright}

\begin{center}
{\Large
Virtual Compton Scattering from the Proton and the  Properties of Nucleon Excited States
}
\end{center}						     

\begin{center}
G.Q. Liu$^*$, A.W. Thomas$^{\dagger}$
\end{center}

\begin{center}
{\it  Department of Physics and Mathematical Physics \\
 University of Adelaide, Adelaide, SA 5005, Australia }
\end{center}

\begin{center}
 P.A.M. Guichon$^\ddagger$
\end{center}

\begin{center}
{\it  DAPNIA/SPhN, CE Saclay, F91191  Gif sur Yvette, France}
\end{center}

\begin{abstract} 
We calculate the $N^*$ contributions to the  
generalized polarizabilities of the proton in virtual Compton 
scattering.  The following nucleon excitations are included: 
 $N^*(1535)$, $N^*(1650)$, $N^*(1520)$, 
$N^*(1700)$, $\Delta(1232)$, $\Delta^*(1620)$ and  $\Delta^*(1700)$.
The relationship between nucleon structure parameters, $N^*$ properties  
and the generalized polarizabilities of the proton is illustrated.
\end{abstract}

\vspace*{0.50cm} 
\noindent -------------------------------------------------------------- \\
$*$ e-mail: gliu@physics.adelaide.edu.au \\
$\dagger$ e-mail: athomas@physics.adelaide.edu.au\\
$\ddagger$ e-mail: pampam@phnx7.saclay.cea.fr

\newpage

\noindent {\bf 1. INTRODUCTION} \\

The study of Virtual Compton Scattering
(VCS), $e+p \longrightarrow e' +p' +\gamma$,
at CEBAF and MAMI (Audit {\it et al.} 1993) could provide valuable information
on the structure of the nucleon, complementing the information obtained from 
elastic form factors, 
real Compton scattering, and deep inelastic scattering. In this paper  we concentrate on 
the kinematic region where the final photon has low energy --- i.e., 
below the threshold for $\pi^0$ production. As shown by 
Guichon {\it et al.} (1995) the low energy cross 
sections are parametrized by 10 generalized polarizabilities 
(GP), functions of the 
virtual photon mass. Their evaluation 
requires the knowledge of the nucleon excited 
states. This sensibility to the nucleon spectrum can provide 
substantial insight into the non-perturbative aspects of the QCD Hamiltonian. 

Guichon {\it et al.} (1995) made an initial evaluation of the GP's 
so as to provide an order of 
magnitude estimate of these new quantities and to illustrate 
their variation as a function of 
the virtual photon mass. In that calculation we neglected 
all recoil effects, that is 
terms which go like the velocity of the nucleon. As a result of that
approximation only 7 GP's were non-zero. 

In this paper we extend the calculations to include the 
recoil corrections which turn 
out to contribute only when the final photon is magnetic. 
We also study the relationship of the nucleon excited states  to  the GP's.
We use the Non-Relativistic Quark Model (NRQM) to take  
advantage of its simplicity and the ready 
availability of its wave functions. Also the separation of the center of mass  and 
internal
motion greatly simplifies the calculation, making it analytically tractable. In 
principle, it is possible to use other wave functions but this can 
be prohibitively laborious and messy.  The NRQM estimate 
should be a useful guide  to the analysis of the soon available experimental data on the  
$p(e,e'p)\gamma$ reaction. \\

\noindent {\bf 2. GENERAL FORMS OF GP'S IN TERMS OF CURRENT DENSITIES} \\

We first briefly outline the formalism for the definition and the calculation 
of the GP's.  We refer to Guichon {\it et al.} 
(1995) for a detailed account of the problem as well as for the 
notations and conventions. 
The hadronic tensor (see Figure 1) is defined by 

\[ H^{\mu\nu}_{NB}(\bq'm_s',\bq m_s) = \int d\bp_X \sum_{X \ne N}
   \left[ \bra N(\bp')|J^{\mu}(0)|X(\bp_X) \ket 
         \frac{\delta(\bp_X-\bp'-\bq')}{E_N(\bq')+q'-E_X(\bp_X)} 
       \bra X(\bp_X)|J^{\nu}(0)|N(\bp) \ket \right.   \]
\begin{equation}
\label{JJ}
   \left. + \bra N(\bp')|J^{\nu}(0)|X(\bp_X) \ket 
         \frac{\delta(\bp_X-\bp+\bq')}{E_N(\bq)-q'-E_X(\bp_X)} 
       \bra X(\bp_X)|J^{\mu}(0)|N(\bp) \ket  \right] + H^{\mu\nu}_{seagull},
\end{equation}
where $J^{\mu}$ is the hadronic current,  $X$ the intermediate baryon excitations 
 and $H^{\mu\nu}_{seagull}$ the contact term generally required by gauge invariance. The reduced 
multipoles are then defined according to
\[ H^{(\rho'L',\rho L)S}_{NB}(q',q) = \frac{1}{2S+1} \sum_{m_s'm_s M'M} 
(-)^{\hf+m_s'+L+M}   \bkk{\hf}{-m_s'}{,\hf}{m_s}{S}{s}
\]
\begin{equation}
\label{RMP}
     \bkk{L'}{M'}{,L}{M}{S}{s} 
     H^{\rho'L'M',\rho L M}_{NB}(q'm_s',qm_s).
\end{equation}
with
\begin{equation}
 H^{\rho'L'M',\rho L M}_{NB}(q'm_s',qm_s) = (4\pi)^{-1}
     \int d\hat{q} d\hat{q'} V^*_{\mu}(\rho' L'M',\hat{q'}) 
     H^{\mu\nu}_{NB}(\bq'm_s',\bq m_s) V_{\nu}(\rho L M,\hat{q}),
\end{equation}
where $V_{\nu}(\rho L M,\hat{q})$ are the charge ($\rho=0$), magnetic ($\rho=1$) and electric 
($\rho=2$) basis vectors defined by Guichon {\it et al.} (1995).

When $\rho ,\ \rho'$ are equal to 0 or 1 the GP's are defined by 
\begin{equation}
P^{(\rho'L',\rho L)S}(q) = \left[ \frac{1}{q'^{L'}q^L} H^{(\rho'L',\rho L)S}_{NB}(q',q) 
\right]_{q'=0}. 
\end{equation}

In the case of a virtual electric photon the analogous definition does not 
yield a GP with the usual photon limit as $q\rightarrow 0$. 
As explained by Guichon {\it et al.} (1995), one must therefore
introduce mixed GP's according to 
\begin{equation}
\hat{H}^{\rho'L'M',L M}_{NB}(q'm_s',qm_s) = (4\pi)^{-1}
    \int d\hat{q} d\hat{q'} V^*_{\mu}(\rho'L'M',\hat{q'}) \sum_i
    H^{\mu i}_{NB}(\bq'm_s',\bq m_s) \left(\cbvy_{M}^{LL+1}(\hat{q})\right)^i, 
\end{equation}
\begin{equation}
\hat{P}^{(\rho'L', L)S}(q) = \left[ \frac{1}{q'^{L'}q^{L+1}} 
     \hat{H}^{(\rho'L', L)S}_{NB}(q',q) \right]_{q'=0}, 
\end{equation}
where $\cbvy_{M}^{Ll}(\hat{q})$ is the vector spherical harmonic. 
The 10 independent GP's needed to describe the low energy regime are then the following,
\[  P^{(11,00)1},  P^{(11,02)1}, P^{(11,11)0}, P^{(11,11)1}, 
\hat{P}^{(11,2)1}, \]
\begin{equation}
 P^{(01,01)0},  P^{(01,01)1}, P^{(01,12)1}, \hat{P}^{(01,1)0}, 
\hat{P}^{(01,1)1}.
\end{equation}
In the low energy regime the following excited states contribute:
$\Xa$, $\Xb$, $\Xc$, $\Xd$, $\Xe$, $\Xf$  and $\Xg$.  

In the NRQM the current density in Eqn.(\ref{JJ}) takes the form 
\[ \bra X(\bp_X)|J^0(0)|N(\bp)\ket = N_0 \bigrho_X(\bp_X-\bp) \],
\begin{eqnarray}
 \bra X(\bp_X)|\bJ(0)|N(\bp)\ket &=& N_0\left[ \frac{\bp_X+\bp}{6m_q}
   \bigrho_X(\bp_X-\bp) +\bP_X(\bp_X-\bp) \right. \nonumber \\
&+& \left. \frac{\bi}{2m_q}  \bSigma_X(\bp_X-\bp)
  \times  (\bp_X-\bp) \right] .
\label{GJ}
\end{eqnarray}
Here $\bigrho, \bP$ and $\bSigma$ are overlap integrals of  
current operators. If one takes into account the factorisation 
of the c.m. and internal baryon wave 
functions, they can be written in the following forms.
\begin{equation}
\label{RHO}
 \bigrho_X(\bp_X-\bp) = \int d\brho d\blambda\, 
   \be ^{-\bi\sqrt{\twotrd}(\bp_{_X}-\bp)\bigcdot\blambda}\,
   \bigphi^{^\dagger}_X(\brho,\blambda, J_X,M_X,T_X,\tau_X)\,
   \hat{Q}\,  \bigphi_N(\brho,\blambda, m_{_N},\tau_{_N}), 
\end{equation}

\[ \bP_X(\bp_X-\bp) = \sqrt{\frac{2}{3}}\frac{1}{2m_q} 
\int d\brho d\blambda\, 
   \be ^{-\bi\sqrt{\twotrd}(\bp_{_X}-\bp)\bigcdot\blambda}\,
\]
\begin{equation}
\label{P}
   \bigphi^{^\dagger}_X(\brho,\blambda, J_X,M_X,T_X,\tau_X)\,
   (\bi \stackrel{\rightarrow}{\boldmath \nabla}_{\lambda} 
  - \bi \stackrel{\leftarrow}{\boldmath \nabla}_{\lambda} )
   \hat{Q}\,    \bigphi_N(\brho,\blambda, m_{_N},\tau_{_N}),
\end{equation}

\begin{equation}
\label{SIGMA}
  \bSigma_X(\bp_X-\bp) = \int d\brho d\blambda\, 
   \be ^{-\bi\sqrt{\twotrd}(\bp_{_X}-\bp)\bigcdot\blambda}\,
   \bigphi^{^\dagger}_X(\brho,\blambda, J_X,M_X,T_X,\tau_X)\,
   \bsigma_3 \hat{Q}\,
   \bigphi_N(\brho,\blambda, m_{_N},\tau_{_N}), 
\end{equation}
where $\hat{Q} = (\frac{1}{6}+\tau_3^z/2)$ is the charge operator 
of the third quark and $\bsigma_3$ is the twice the spin operator of the third quark.

With specific internal wave functions for the nucleon ($\bigphi_N$) and the 
intermediate excitations ($\bigphi_X$), one obtains 
explicit forms for  the current density and hence the hadronic tensor. 
\\

\noindent {\bf 3. CURRENT DENSITY AND HADRONIC TENSOR IN NRQM} \\

To calculate the  integrals $\bigrho(\bp_X-\bp)$, $\bP(\bp_X-\bp)$,  
and $\bSigma(\bp_X-\bp)$  for intermediate states $X$ being
$\Xa$, $\Xb$, $\Xc$, $\Xd$, $\Xe$, $\Xf$  and $\Xg$, 
we use the wave functions from Isgur and Karl (1978).  
The calculation is  straightforward though fairly
lengthy. We arrive at the following expressions.

\begin{equation}
\label{RHO2}
 \bigrho_{_{N(^28)}}(\bq)  = -\bi \frac{\sqrt{8\pi}}{9}  
    \frac{q}{\alpha} \expfac  \tau_{_N} 
    \trija{1}{M_X\!\!-\!m_{_N}}{\hf}{m_{_N}}{J_X}{M_X} 
    Y_{1M_X-m_{_N}}^*(\hat{q}).
\end{equation}

\begin{equation}
  \bP_{_{N(^28)}}(\bq) = -\bi \frac{\sqrt{6}}{9} 
    \frac{\alpha}{m_q} \expfac  \tau_{_N} 
    \trija{1}{M_X\!\!-\!m_{_N}}{\hf}{m_{_N}}{J_X}{M_X} 
   \bolde^*_{M_X-m_{_N}}.
\end{equation}

\[
\bSigma_{_{N(^28)}}(\bq) = -\bi \frac{\sqrt{\pi}}{18} 
    \frac{q}{\alpha} \expfac 
   (1+4\tau_{_N}) 
   \sum_{\mu} 
    \trija{1}{M_X\!\!-\!\mu}{\hf}{\mu}{J_X}{M_X}
\]
\begin{equation}
   \hspace*{8.0cm} Y^*_{1M_X-\mu}(\hat{q}) \cb{\mu}
\end{equation}

\[  \bSigma_{_{N(^48)}}(\bq) = -\bi \frac{\sqrt{\pi}}{18} 
    \frac{q}{\alpha} \expfac 
  (1-2\tau_{_N})  
   \sum_{\mu} 
 \trijb{1}{M_X\!\!-\!\mu}{\thf}{\mu}{J_X}{M_X} 
\]
\begin{equation} 
  \hspace*{8.0cm} Y^*_{1M_X-\mu}(\hat{q})  \ca{\thf}{\mu}
\end{equation}

\begin{equation}
 \bSigma_{_{\Delta}}(\bq) = \frac{1}{3} \expfac 
   \ca{\thf}{m_{\Delta}} 
\end{equation}

\begin{equation} 
\bigrho_{_{\Delta^{^*}}}(\bq) = \frac{1}{2\tau_{_N}} \bigrho_{_{N(^28)}}(\bq)
\end{equation}

\begin{equation} 
\bP_{_{\Delta^{^*}}}(\bq) = \frac{1}{2\tau_{_N}} \bP_{_{N(^28)}}(\bq)
\end{equation}

\begin{equation} 
\label{SIGMA2}
\bSigma_{_{\Delta^{^*}}}(\bq)=\frac{-1}{1+4\tau_{_N}} \bSigma_{_{N(^28)}}(\bq)
\end{equation}

where $\chi$ are spin wave functions,   $\tau_{_N}$ the isospin quantum number 
of the nucleon (i.e., $\pm\hf$ for $(^p_n)$), and 
$\bolde_{m}$ is the $m$th component of the spherical basis vectors.
Eqns.(\ref{RHO2}-\ref{SIGMA2}) are for intermediate states with 
the same isospin  quantum number as the proton.

The main characteristics of these integrals are summarized below: 

a.) The $\Delta(1232)$ and the $^48$  component of the $N^*$'s contribute only
to $\bSigma(\bp_X-\bp)$ .

b.) The $^28$  component of the $N^*$'s, and  the $\Delta^*$'s  contribute
to all $\bigrho(\bp_X-\bp)$, $\bP(\bp_X-\bp)$,  and $\bSigma(\bp_X-\bp)$.

c.) For small $x=|\bp_X-\bp|$ the behavior of the  integrals are
$ \bSigma_{\Delta}(\bx), \bP(\bx) \propto  O(1),$ and 
$\bSigma_{N^*,\Delta^*}(\bx),\bigrho(\bx) \propto  O(x).$

The leading term and the recoil term of the hadronic tensor in
 Eqn.(\ref{JJ}) are separated  in the following way.
We work in the initial $N\gamma$ c.m. system so that $\bp=-\bq$.
In the direct term of Eqn.(\ref{JJ}), $\bp_X=0$ and $\bq'=-\bp'$ . 
So, aside from the energy denominator, the direct amplitude 
factorizes into a product of a $\bq$--dependent current density 
$ \bJ_{d,XN}(\bq) $ and a $\bq'$ dependent current density 
$ \bJ_{d,NX}(\bq')$, where

\[ \bJ_{d,XN}(\bq) 
 = \bra X(\bp_X)|\bJ(0)|N(\bp)\ket_{direct} 
 = N_0\left[ \frac{-\bq}{6m_q}
   \bigrho_X(\bq) +\bP_X(\bq) 
+ \frac{\bi}{2m_q}  \bSigma_X(\bq)  \times  \bq \right],
\]

\[
 \bJ_{d,NX}(\bq')  
 =\bra N(\bp')|\bJ(0)|X(\bp_X)\ket_{direct}  
 =N_0\left[ \frac{-\bq'}{6m_q}
   \bigrho_X^*(\bq') +\bP_X^*(\bq') 
- \frac{\bi}{2m_q}  \bSigma_X^*(\bq')  \times  \bq' \right].
\]

The current density in the cross term has a more complicated $\bq$ and
$\bq'$ dependence.  
Because  $\bp_X=-\bq-\bq'$, $\bp+\bp_X = -\bq'-2\bq$ and 
$\bp+\bp_X' =-\bq-2\bq'$, it involves terms depending upon both 
$\bq$ and $\bq'$.  Let 
\[
 \bra X(\bp_X)|\bJ(0)|N(\bp)\ket_{cross} =  \bJ_{c,XN}(\bq') 
  + \delta\bJ_{XN}(\bq, \bq'),
\]
and 
\[
 \bra N(\bp')|\bJ(0)|X(\bp_X)\ket_{cross} =  \bJ_{c,NX}(\bq) +
 + \delta\bJ_{NX}(\bq, \bq'),
\]
then   $ \bJ_{c,XN}(\bq')$ , $\bJ_{c,NX}(\bq)$, $\delta\bJ_{XN}(\bq, \bq')$
and $\delta\bJ_{NX}(\bq, \bq')$ are given by the following expressions:
\[
 \bJ_{c,XN}(\bq')=N_0\left[ \frac{-\bq'}{6m_q} \bigrho_X(-\bq') +\bP_X(-\bq') 
 + \frac{\bi}{2m_q}  \bSigma_X(-\bq')  \times  (-\bq') \right],
\]
\[
 \bJ_{c,NX}(\bq)=N_0\left[\frac{-\bq}{6m_q}\bigrho_X^*(-\bq)+\bP_X^*(-\bq) 
  - \frac{\bi}{2m_q}  \bSigma_X^*(-\bq)  \times  (-\bq) \right],
\]
\[
 \delta\bJ_{XN}(\bq, \bq') =  N_0 \frac{-\bq}{3m_q} \bigrho_X(-\bq'),   
\]
\[
 \delta\bJ_{NX}(\bq, \bq') = N_0 \frac{-\bq'}{3m_q} \bigrho_X^*(-\bq).
\]

From Eqn.(\ref{GJ}) we see that $J^0$ depends only on $\bq$ or $\bq'$,  
so that  $\delta J^{0} =0$. 
We define the {\it leading term} of the hadronic tensor 
$H^{\mu\nu}_{NB}(\bq'm_s',\bq m_s)$  by neglecting the terms 
depending on both $\bq$ and $\bq'$ in the cross term, and  the 
 $\bq'$ dependence of $E_X(\bq+\bq')$ in the energy denominator 
of the cross amplitude. That is, 
\begin{equation}
\label{JJLEAD}
 H^{\mu\nu}_{NB-LEADING}(\bq'm_s',\bq m_s) = \sum_{X \ne N} \left[
     \frac{ J^{\mu}_{d,NX}(\bq')\, J^{\nu}_{d,XN}(\bq) }{\dMx}
   + \frac{ J^{\mu}_{c,XN}(\bq')\, J^{\nu}_{c,NX}(\bq) }{\dEx}
    \right]+ H^{\mu\nu}_{seagull}. 
\end{equation}

The {\it recoil term} of $H^{\mu\nu}_{NB}(\bq'm_s',\bq m_s)$  
includes all the effects coming 
from terms depending on both $\bq$, and $\bq'$, and those  arising from the 
expansion of the energy denominator of the cross term to order $q'$.
The recoil contribution contains terms to order $q'$ and higher and 
can be written as follows,
\[ H^{\mu\nu}_{NB-RECOIL}(\bq'm_s',\bq m_s) = \sum_{X \ne N} \left[
  \frac{   J^{\mu}_{c,XN}(\bq')\,\delta J^{\nu}_{NX}(\bq, \bq') 
   + \delta J^{\mu}_{XN}(\bq,\bq')\, J^{\nu}_{c,NX}(\bq)}{\dEx} \right.
\]
\begin{equation}
\label{JJRECOIL}
\left. + \frac{1}{[\dEx]^2} \left(q'+\frac{\bq'\cdot\bq}{E_X(q)} \right) 
  J^{\mu}_{c,XN}(\bq')\,J^{\nu}_{c,NX}(\bq) \right].
\end{equation}
To get the GP we need only  keep 
terms to order $q'$. So the recoil effects contribute only
to GP's with a magnetic final photon, $\mu\ne 0$. 

With Eqns.(\ref{JJLEAD},\ref{JJRECOIL}) for the 
{\it leading } and {\it recoil terms}, one can carry out the 
partial wave projection using the definitions of section 2. The partial wave
projection poses no particular difficulty except the need for 
careful book-keeping. We therefore skip the details of the partial 
wave decomposition and give the final expressions for the GP's
in section 4.\\ 

\noindent {\bf 4. GENERALIZED POLARIZABILITIES IN THE NRQM}\\

Here we give the final results for the 10 GP's in the NRQM. 
We also  study the  properties of the 
GP's in relation to the 
parameters determining the nucleon structure in the NRQM.
Our aim is  to find those properties of the nucleon structure 
to which the GP's are most sensitive.  
This should help identify the most useful 
aspects of VCS in studying the nucleon structure.

The analytic expressions for the 10 GP's can be written 
in the following form, where the leading contributions 
are the same as  in Guichon {\it et al.} (1995). (Note, however,
that the curve for $P^{(01,1)S}$ in Guichon {\it et al.} has an error where
the cross term was a factor of 2 too large. We correct it here.
The overall sign for $P^{(01,1)S}$ is also changed to conform with
the definition for the  electric virtual photon case.)

The GP's are plotted  in Figure 2. \\

\noindent {\bf 4.1  Leading Contributions To The Generalized Polarizabilities }

\begin{equation}
P^{(0101)S} = {1 \over 2S+1} {1 \over 18} {1 \over \alpha^2}\expfac  
       \sumX a_X^2   \left( {Z_d^{S,J_X} \over \dMx } 
      + {Z_c^{S,J_X} \over \dEx} \right) \hspace*{1.20cm} 
\end{equation}

\begin{equation}P^{(0112)1} = {1 \over 108} \sqrt{3 \over 5}  
   {1 \over m_q\alpha^2} \expfac  
     \sumX  a_X^2 {(-1)^{I_x-1/2} \over 2I_x}
       \left( {  Z_{ad}^{2,S,J_X} \over \dMx } -
     {  Z_{ac}^{2,S,J_X}   \over  \dEx} \right) 
\end{equation}

\begin{equation}
P^{(1111)S}_{para}={1 \over 2S+1}{4 \over 27}{1 \over m_q^2}\expfac
  \left({ Z_\Delta^S\over M-M_\Delta }+{ Z_\Delta^S\over E(q)-E_\Delta(q)} \right) 
\hspace*{3.70cm}
\end{equation}

\begin{equation}
P^{(1111)S}_{dia} = \delta_{S0}{7\sqrt{6} \over 54} 
      {1 \over m_q\alpha^2}\expfac  \hspace*{8.50cm}
\end{equation}
\noindent The mixed GP is the sum of two terms:
\begin{equation}
\hat{P}^{(01,1)S}= \hat{P}^{(01,1)S}_F + \hat{P}^{(01,1)S}_S \hspace*{9.0cm}
\end{equation}
\begin{equation}
\hat{P}^{(01,1)S}_F = {1 \over 2S+1} {1 \over 108} 
    {1 \over m_q\alpha^2}\expfac \sumX a_X^2 
      \left( {Z_d^{S,J_X} \over \dMx } 
      - {Z_c^{S,J_X} \over \dEx} \right)\hspace*{0.0cm}
\end{equation}

\begin{equation} 
\hat{P}^{(01,1)S}_S ={1 \over 2S+1}{1 \over 36\sqrt{3}}
   {1 \over m_q\alpha^2} \expfac  
   \sumX a_X^2 {(-1)^{I_x-1/2} \over 2I_x}
    \left({ Z_{ad}^{1,S,J_X}  \over \dMx } 
       - {Z_{ac}^{1,S,J_X}  \over \dEx }\right)
\end{equation}

Angular functions  $Z^{S,J_X}$, $Z^{L,S,J_X}$ and $Z^S_{\Delta_{1232}}$ 
in the above summation are given by:

\def\tvi{\vrule height 12pt depth 5pt width 0pt}
\def\tv{\tvi\vrule}

$$\vbox{\offinterlineskip
\halign{\tv\hfil\enskip#\enskip\hfil\tv&\hfil\enskip
#\enskip\hfil\tv&\hfil\enskip#\enskip\hfil\tv
&\hfil\enskip#\enskip\hfil\tv&\hfil\enskip#\enskip\hfil\tv
&\hfil\enskip#\enskip\hfil\tv
&\hfil\enskip#\enskip\hfil\tv&\hfil\enskip#\enskip\hfil\tv\cr
 \noalign{\hrule}
L & S & $J_X$  &  $Z_d^{S,J_X}$  &  $Z_c^{S,J_X}$  & $Z_{ad}^{L,S,J_X}$      
  & $Z_{ac}^{L,S,J_X}$ & $Z_{\Delta_{1232}}^{S}$  \cr\noalign{\hrule}
1 & 0 & 1/2  & $\sqrt{2/3}$ & $\sqrt{2/3}$ & $-2/\sqrt{3}$ 
  & $2/\sqrt{3}$  &  \cr  \noalign{\hrule}\tvi
1 & 0 & $3/2$ & $2\sqrt{2/3}$ & $2\sqrt{2/3}$ & $2/\sqrt{3}$ & $-2/\sqrt{3}$ 
  & $\sqrt{6}$   \cr\noalign{\hrule}\tvi 
1 & 1 & 1/2  & 2    & $-2/3$   & $-2\sqrt{2}$   &$-2\sqrt{2}/3$ 
  &  \cr\noalign{\hrule}\tvi 
1 & 1 & 3/2 & $-2$    & 2/3   & $-\sqrt{2}$   &$-\sqrt{2}/3$ 
  & $1$ \cr\noalign{\hrule}\tvi 
2 & 1 & 3/2 &     &       & $\sqrt{30}$  &$\sqrt{30}/3$  & 
\cr\noalign{\hrule} 
}}$$

\noindent {\bf 4.2 Recoil  Contributions To Generalized Polarizabilities }

\begin{equation}
P^{(1100)1}_{recoil} = -  {1 \over 3\sqrt{3}\, }
      {q^2  \over  m_q } \,\expfac
       \sumX {a_X^2 Z_{1100}^{J_X} \over E_X(q) [\dEx]^2 }  
       \left[1-  {E_X(q) [\dEx] \over 3\alpha^2 }\right]
\end{equation} 

\begin{equation}
P^{(1102)1}_{recoil} =   - {1 \over 3\sqrt{3}\,} {1\over m_q }\,\expfac
       \sumX {a_X^2 Z_{1102}^{J_X}\over E_X(q) [\dEx]^2 }  
       \left[1-  {E_X(q) [\dEx] \over 3\alpha^2 }\right] 
\end{equation} 

\begin{equation}
 P^{(1111)0}_{recoil} =  {1 \over 3\, } {\alpha^2 \over  m_q^2 } \,\expfac
       \sumX { a_X^2 Z_{1111}^{J_X} \over E_X(q) [\dEx]^2 } 
    \hspace*{0.0cm}
\end{equation} 

\begin{equation}
\hat{P}^{(112)1}_{recoil} = - {1 \over 6\sqrt{15}\,}{1 \over m_q^2 }\,\expfac
       \sumX {a_X^2  Z_{1102}^{J_X} \over E_X(q) [\dEx]^2 }  
       \left[1-  {E_X(q) [\dEx] \over 3\alpha^2 }\right] 
\end{equation} 
where

$$\vbox{\offinterlineskip
\halign{\tv\hfil\enskip#\enskip\hfil\tv&\hfil\enskip
#\enskip\hfil\tv&\hfil\enskip#\enskip\hfil\tv
&\hfil\enskip#\enskip\hfil\tv\cr
 \noalign{\hrule}
$J_X$ &  $Z^{J_X}_{1100}$ & $Z^{J_X}_{1102}$  & $Z^{J_X}_{1111}$\cr\noalign{\hrule}
  1/2 & $2/27$ &  $\sqrt{2}/27$ & $3\sqrt{6}/27$ \cr\noalign{\hrule}\tvi
  3/2 & $-2/27$ &  $-\sqrt{2}/27$ & $6\sqrt{6}/27$ \cr \noalign{\hrule}
}}$$

The $^48$ component of the excited state wavefunctions 
contribute nothing to the proton GP's, because of the isospin 
factor $(1-2\tau)$. However, they do contribute in the neutron case,
which we do not study here. The parameters $m_q=350$  MeV and 
$\alpha=320$ MeV are 
taken from Isgur and Karl (1978). The P-wave intermediate states are ordered as 
in the following table, together with their representation mixing 
parameters $a_X$ for the $^28$ representation.

$$\hspace*{-0.8cm}\vbox{\offinterlineskip
\halign{\tv\hfil\enskip#\enskip\hfil\tv&\hfil\enskip
#\enskip\hfil\tv&\hfil\enskip#\enskip\hfil\tv
&\hfil\enskip#\enskip\hfil\tv&\hfil\enskip#\enskip\hfil\tv
&\hfil\enskip#\enskip\hfil\tv&\hfil\enskip#\enskip\hfil\tv\cr
 \noalign{\hrule}
 X&$\Xa$ & $\Xb$ &$\Xc$& $\Xd$ & $\Xf$ & $\Xg$\cr\noalign{\hrule}
$a_X$  & $0.85$ & $-0.53$ & $0.99$ & $0.11$ & $1.0$ & $1.0$
\cr\noalign{\hrule} 
 }}$$

\newpage
\noindent {\bf 5. $N^*$ PROPERTIES AND THE GP'S}\\

The NRQM  parameters are well determined  by fitting the static properties
of baryons (Isgur and Karl, 1978). 
Nonetheless, there are other phenomenological
models than the NRQM, such as the bag models.  Different models may 
not necessarily give exactly the same properties for the nucleon 
and its excitations. We do not intend to survey all the different nucleon 
models in this paper, but just to investigate those nucleon and $N^*$ 
properties which exert the most important influence on the behavior of 
the GP's in the NRQM. Two main factors are studied here: 
the mass spectrum of the nucleon 
and the size parameter $\alpha$. The results are displayed in 
Figures 2, 3, and 4.

 The GP's have a strong dependence on the mass and energy spectrum of the 
excited states of the nucleon and the $\Delta$.
In Figure 2 we used the average masses of the $N^*$ and 
$\Delta^*$ from the particle data tables (1994).  
However, these masses are 
all determined within a range and may be different from the 
predictions of the NRQM.  We study  the effects of 
the  $N^*$ mass spectrum by comparing the GP's calculated
with  the lower and upper limits of the masses from the particle data
tables (1994)
and also with those predicted in NRQM of Isgur and Karl (1978).
Figure 3  shows that some  GP's  are  very  sensitive to the $N^*$ 
masses, particularly at small $q$. The effect on $\hat{P}^{(01,1)0}$ 
is especially large 
as compared with the theoretical masses of Isgur and Karl (1978).  
The change in  $P^{(11,00)1}$, $P^{(11,02)1}$, and $\hat{P}^{(11,2)1}$ are
quite drastic due to a more complicated factor of the mass 
and energy differences.

 The effects of the hadron size parameter $\alpha$, are illustrated 
in Figure 4, where the plot is for a variation of $\pm 5\% $ in 
$\alpha$ from its normal value of 320 MeV. The main influences are 
again seen to be in the low $q$ region.

In conclusion, we have presented a calculation of the $N^*$ contribution 
to the generalized polarizabilities
for virtual Compton scattering on the proton. The dependence of these
GP's on the $N^*$ properties has been studied.  
We hope that there will soon be experimental data with which these 
estimates can be compared.

This work was supported by the Australian Research Council.


\newpage 
\begin{center}
{\bf FIGURE CAPTIONS}
\end{center}

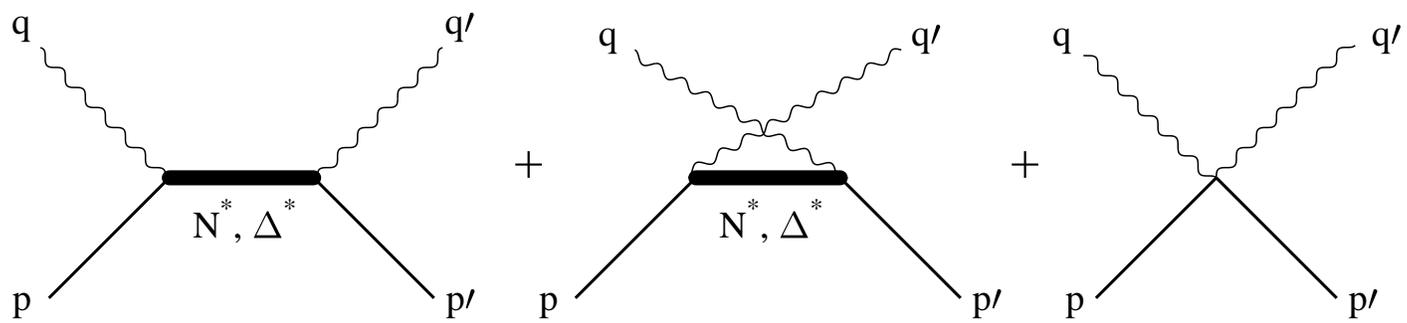
Figure.1. Direct, Cross  and Seagull Terms of the hadronic tensor in the 
lowest order QED perturbation theory.\\

Figure 2. The 10 GP's in the NRQM, The parameters $m_q=350$ MeV 
and $\alpha=320$ MeV. 
Note that  $P^{(11,00)1}$, $P^{(11,02)1}$, and $\hat{P}^{(11,2)1}$ 
are all zero in the absence of the recoil correction. (Note that in Fig.
2c the superscript ``para" refers to the paramagnetic contribution from
the $\Delta(1232)$ and ``dia'' to the seagull contribution which has
the opposite sign.)\\

Figures 3. Effects of the masses of nucleon($\Delta$) excitations on GP's.
Four groups of different masses for P-wave excitations are used. \\ 
dotted line -- using the lower limit of the masses from the particle
data tables (1994): 
$N^*(1520)$ $N^*(1640)$ $N^*(1515)$ $N^*(1650)$ $\Delta^*(1615)$ $\Delta^*(1670)$ \\
dashed line -- using the upper limit of the masses from 
the particle data tables (1994): 
$N^*(1555)$ $N^*(1680)$ $N^*(1530)$ $N^*(1750)$ $\Delta^*(1675)$ $\Delta^*(1770)$ \\
dot-dashed line -- using the theoretical masses given in 
Isgur and Karl (1978):
$N^*(1490)$ $N^*(1655)$ $N^*(1535)$ $N^*(1745)$ $\Delta^*(1685)$ $\Delta^*(1685)$ \\
solid line -- same as Figs.2,using the average masses from  
the particle data tables (1994):
$N^*(1535)$ $N^*(1650)$ $N^*(1520)$ $N^*(1700)$ $\Delta^*(1620)$ $\Delta^*(1700)$ \\

Figures 4.  Effects of a  $\pm 5\% $ variation (304 and 336 MeV) in $\alpha$ from its 
normal value of 320 MeV. Other parameters remain the same as in Figs.2. 
The dotted line is for $\alpha=304$ MeV, the dashed line is for 
$\alpha=336$ MeV, the solid line  $\alpha=320$ MeV, as in Figure 2.\\

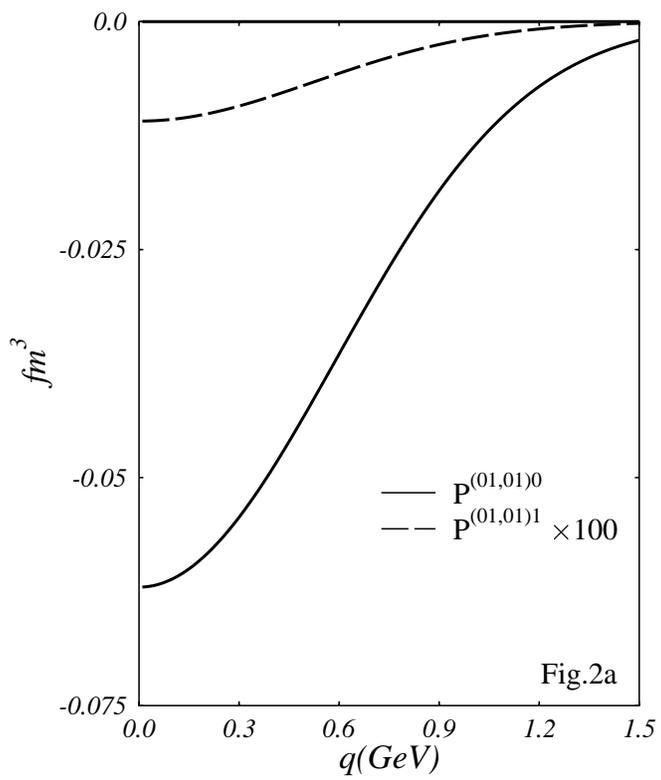
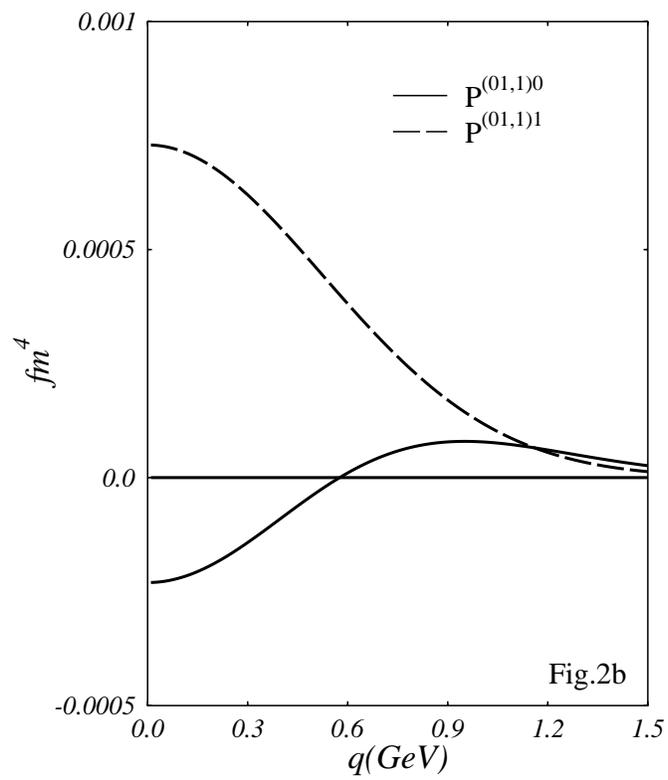
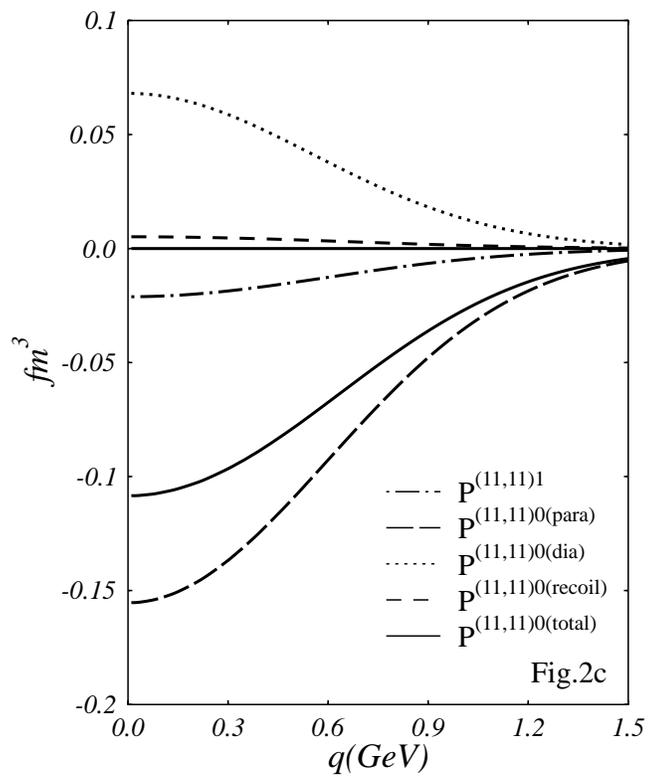
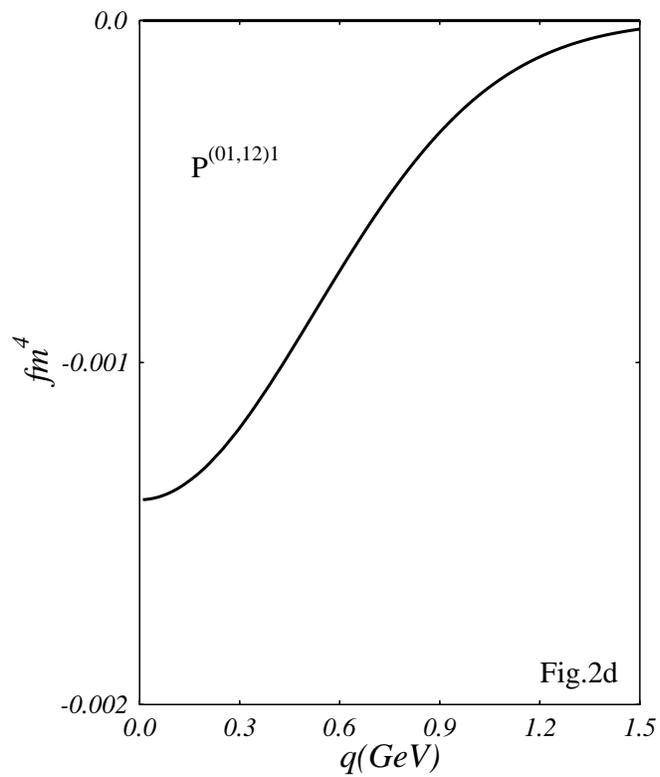
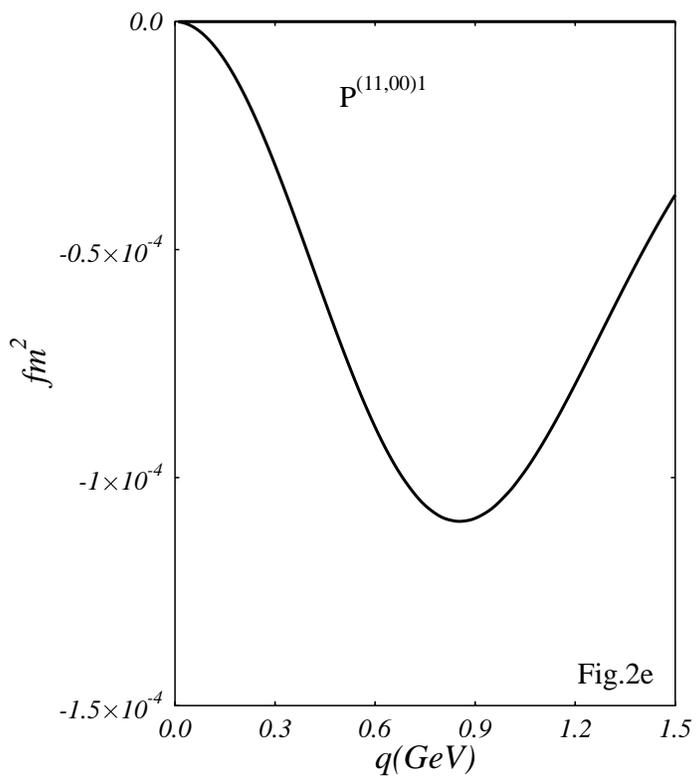
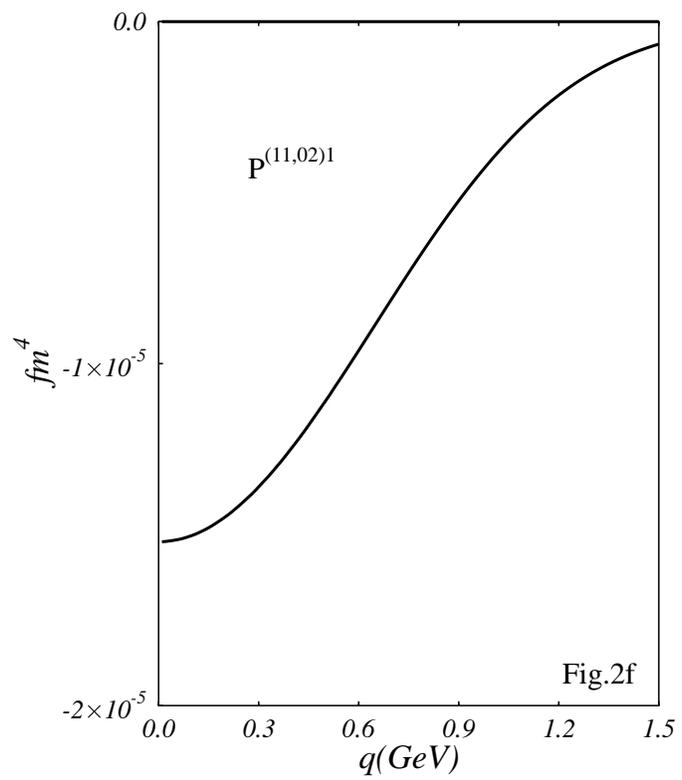
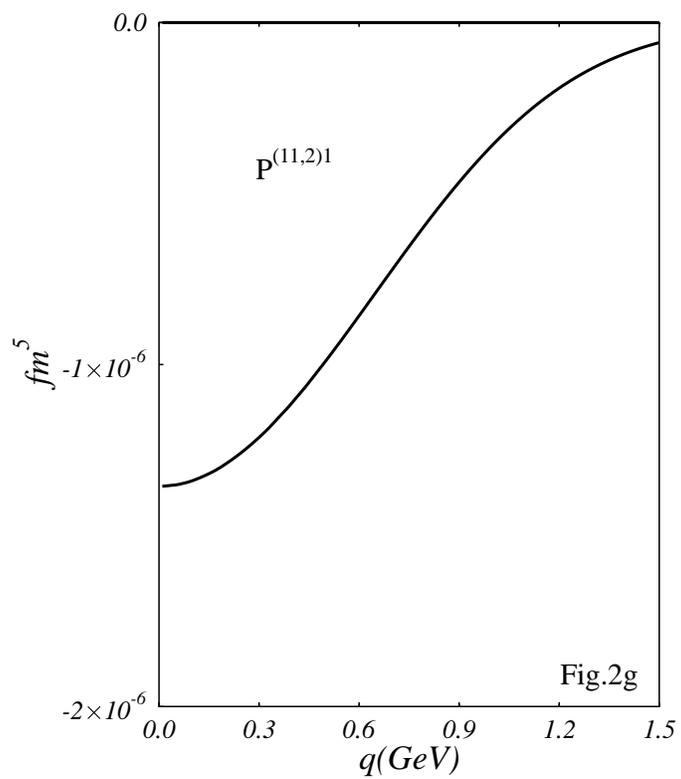
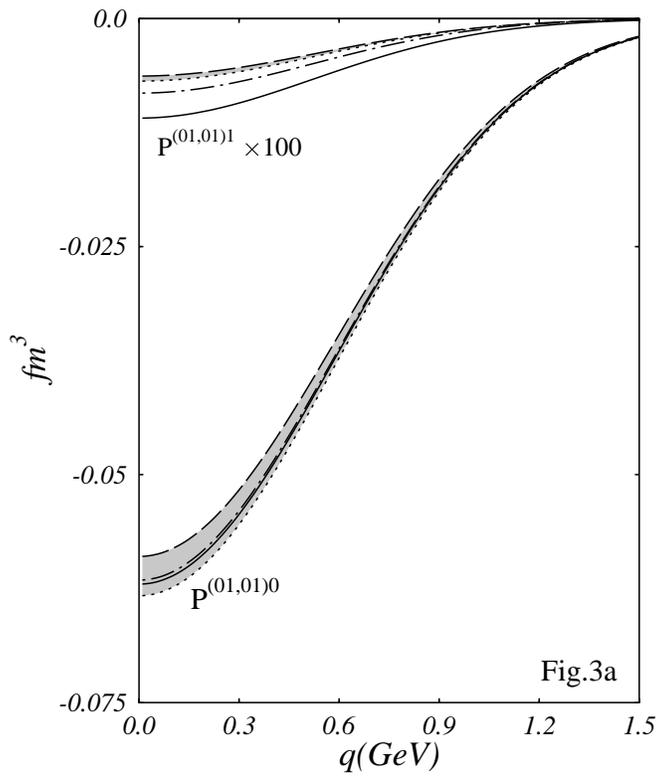
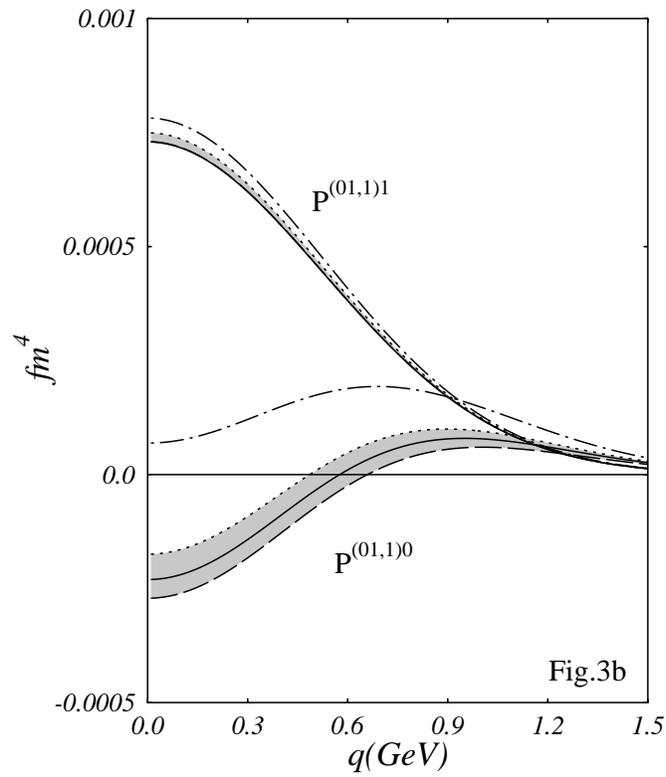
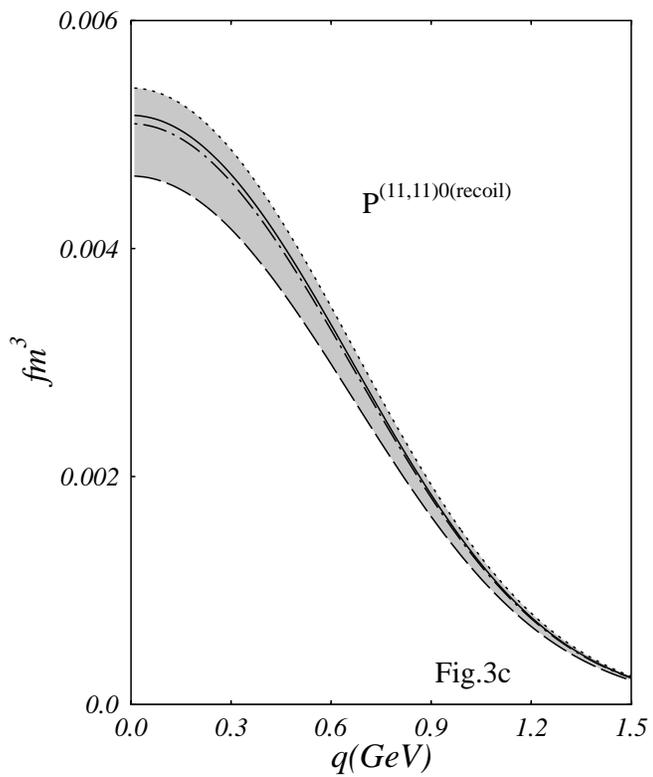
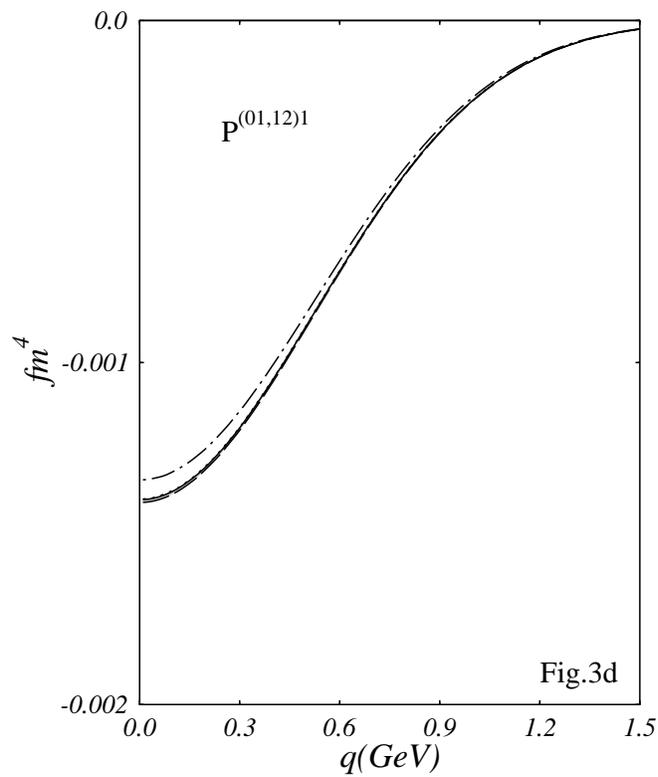
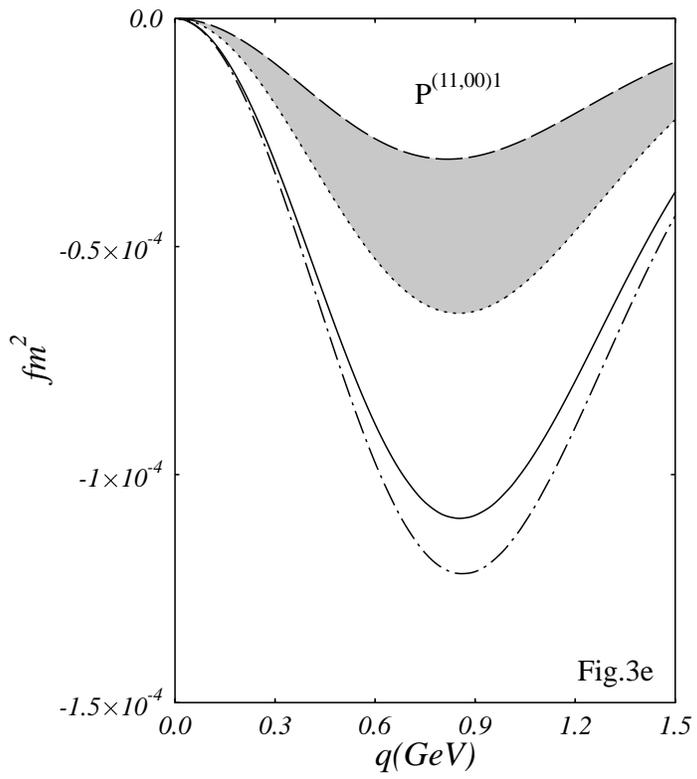
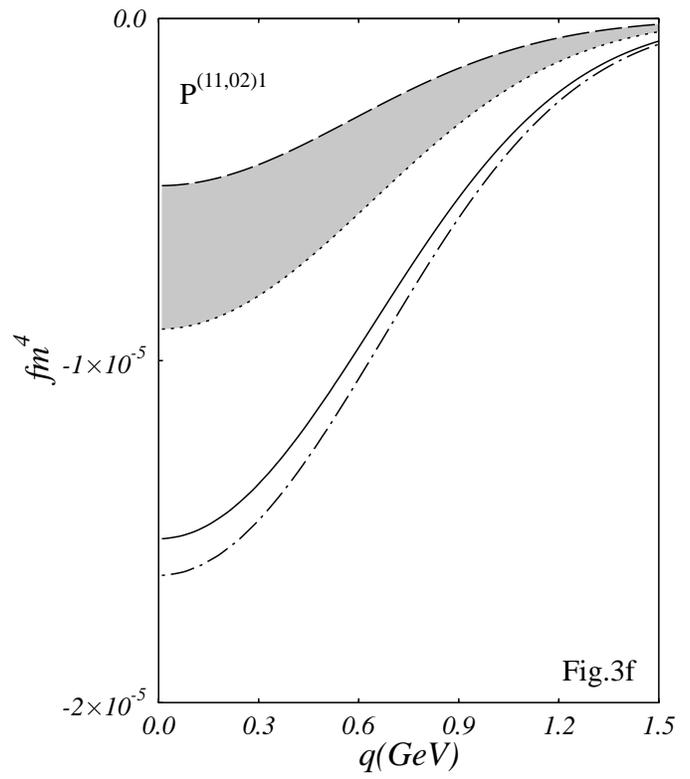
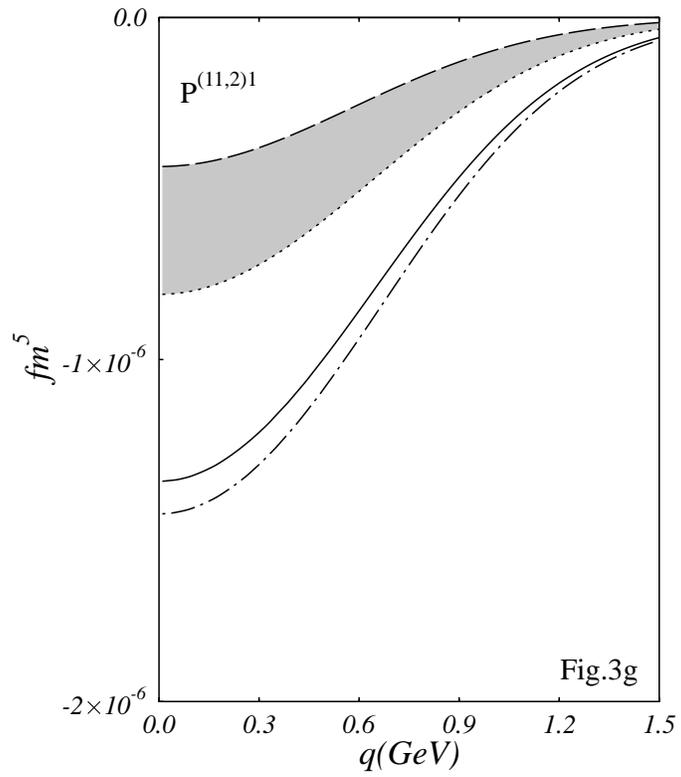
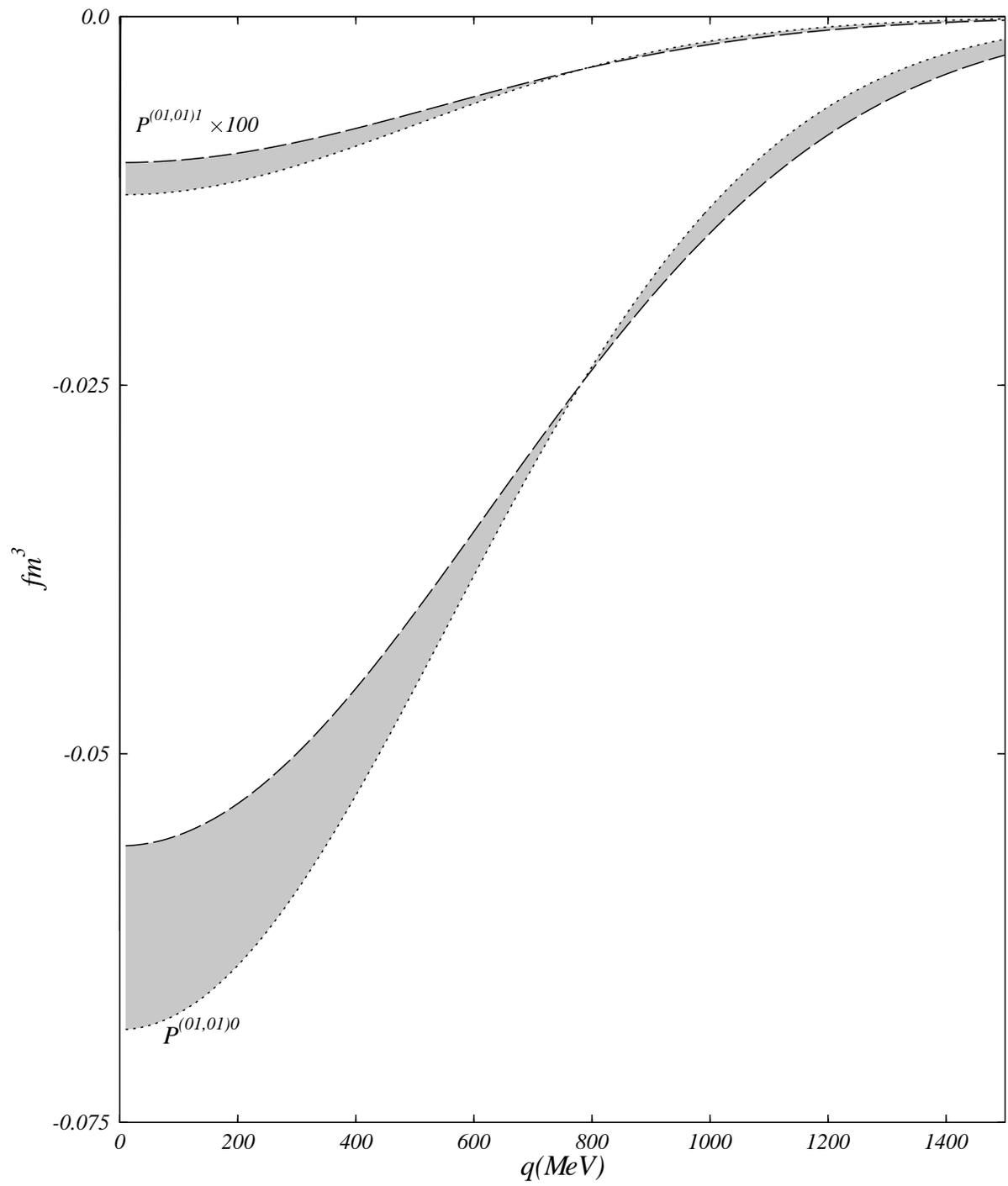
\end{document}